\documentstyle[12pt]{article}
\input epsf
\begin{document}

\newcommand{\eq}{\begin{equation}}
\newcommand{\en}{\end{equation}}
\newcommand{\bino}{\tilde{b}}
\newcommand{\tsquark}{\tilde{t}}
\newcommand{\gluino}{\tilde{g}}
\newcommand{\wino}{\tilde{w}}
\newcommand{\mtilde}{\tilde{m}}
\newcommand{\higgsino}{\tilde{h}}
\newcommand{\gsi}{\,\raisebox{-0.13cm}{$\stackrel{\textstyle>}
{\textstyle\sim}$}\,}
\newcommand{\lsi}{\,\raisebox{-0.13cm}{$\stackrel{\textstyle<}
{\textstyle\sim}$}\,}

\rightline{RU-94-38}
\rightline{hep-ph/9410401}
\rightline{\today}
\baselineskip=18pt
\vskip 0.5in
\begin{center}
{\bf \LARGE Radiative Gaugino Masses}\\
\vspace*{0.4in}
{\large Glennys R. Farrar}\footnote{Research supported
in part by NSF-PHY-91-21039} \\
\vspace{.05in}
{\it Department of Physics and Astronomy \\ Rutgers University,
Piscataway, NJ 08855, USA}\\
\vspace*{0.2in}
{\large Antonio Masiero}\\
\vspace{.04in}
{\it INFN - Sez. di Padova \\  Via Marzolo, 8 -- 35131 Padova, Italy}
\end{center}
\vskip  0.1in

{\bf Abstract:}
We investigate the possibility that gauginos are massless at tree
level and that the U(1) R-invariance is broken spontaneously by Higgs
vevs, like the chiral symmetry of quarks in the standard model, or
else explicitly by dimension 2 or 3 SUSY-breaking terms in the low
energy effective Lagrangian.  Gluino and lightest neutralino masses
then depend on only a few parameters.  For a SUSY-breaking scale $\lsi
400$ GeV, the gluino and lightest neutralino have masses typically in
the range $1/10 \sim 2 \frac{1}{2}$ GeV.  On the other hand, for a
SUSY-breaking scale several TeV or larger, radiative contributions can
yield gluino and lightest neutralino masses of $O(50-300)$ GeV and
$O(10-30)$ GeV, respectively.  As long as the Higgs vev is the only
source of R-invariance breaking, or if SUSY breaking only appears in
dimension 2 terms in the effective Lagrangian, the gluino is
generically the lightest SUSY particle, modifying the usual
phenomenology in interesting ways.

\thispagestyle{empty}
\newpage
\addtocounter{page}{-1}
\newpage

\section{Introduction}
\hspace*{2em}

There is nowadays an intense effort to understand the nature and
structure of the supersymmetry breaking sector in low energy effective
theories which are obtained as the pointlike limit of superstrings.
The pattern of these soft breaking terms is obviously linked to the
mechanism which is chosen in superstrings to originate the breaking of
the local supersymmetry.  One interesting type of SUSY breaking
predicts vanishing gaugino masses at the scale of supergravity
breaking.  This class of superstring models is often discarded on the
phenomenological basis that gaugino masses (in particular the gluino
mass) in the low-energy theory would be too small.  In this paper we
discuss this possibility and we show that scenarios with vanishing
tree-level gaugino masses are not so strictly excluded as is
commonly believed.

R-invariance is automatically a symmetry of the MSSM Lagrangian before
supersymmetry is broken.  In superfield form, the F-terms of this
Lagrangian have the trilinears which are needed to give ordinary
fermions their masses: $( \hat{Q} \hat{U}^c \hat{H}_u)_{\theta
\theta}$ and $( \hat{Q} \hat{D}^c \hat{H}_d)_{\theta \theta}$ and the
analogs for the leptons.  In addition, the term $\mu (\hat{H_u}
\hat{H_d})_{\theta \theta}$ is needed to break the ew gauge symmetry,
at least in the scenario of refs. \cite{a-gpw,ir}.  Assigning
$R(\theta)=1$, an R-charge assignment for the chiral superfields can
be found which satisfies the conditions
\eq
\begin{array}{c}
R(\hat{H_u}) + R(\hat{H_d}) = 2 \\
R(\hat{H_u}) + R(\hat{Q}) + R(\hat{U^c}) = 2 \\
R(\hat{H_d}) + R(\hat{Q}) + R(\hat{D^c}) = 2 \\
R(\hat{H_d}) + R(\hat{L}) + R(\hat{E^c}) = 2
\label{R}
\end{array}
\en
so that the Lagrangian is R-invariant\footnote{See ref. \cite{f:39}
for a more detailed discussion, and the discussion of the vector
superfields.}.  Some soft-SUSY-breaking terms which may be present in
the full Lagrangian break R-invariance, and others do not.
Scalar masses and self-interactions involving $\phi^*
\phi$ are invariant for any choice of the R-charge of the associated
superfield.  However R-invariance for the terms $A \mtilde \tilde{t}
\tilde{t}^c H_u$ and $B \mtilde \mu H_u H_d$ are inevitably
inconsistent with the conditions (\ref{R}), since the $\theta \theta$
which must be factored out in going from the superpotential to the
Lagrangian written in terms of component fields carries R=2.  Thus if
either $A$ or $B$ is non-zero, R-invariance is broken
explicitly\footnote{If one chooses to define R-invariance to be the
chiral symmetry associated with a massless gluino, with no reference
to the transformation of $\theta$ in the superfield, one would still
arrive at the same conclusions, as a result of needing to give
non-trivial transformations to quarks and squarks on account of their
Yukawa couplings to gluinos and to Higgs and higgsinos on account of
their Yukawa couplings to quarks and squarks.}.  However even if
$A=B=0$, R-invariance is broken spontaneously when $<H_u>$ and $<H_d>$
are non-zero\footnote{It is possible to find a solution to the
conditions (\ref{R}) such that either $H_u$ or $H_d$ has $R=0$, but
not both, so that if only one of the Higgs got a vev, R-invariance
would not be broken spontaneously.  Then it would be hard to
understand ordinary fermion masses so we discard this as an option.},
so that gaugino mass terms can be generated radiatively. It is
interesting to consider several different possibilities, always
taking tree-level gaugino masses to be zero:
\begin{enumerate}
\item $A=B=0$.  This corresponds to the possibility that R-invariance
is only broken spontaneously, along with electroweak gauge invariance,
by vevs of the Higgs fields.
\item $A=0$.  This corresponds to the absence of dimension-3
SUSY-breaking terms in the low energy Lagrangian, which arises
naturally in hidden sector models without gauge singlets\cite{bkn}.
\item Non-zero $A$ and $B$.  We consider this for completeness,
in case someday a SUSY-breaking mechanism is discovered
which has this feature.
\end{enumerate}

Years ago, the possibility of tree-level-vanishing gaugino masses in
N=1 supergravity theories was discussed in refs. \cite{bgm}(BGM) and
\cite{bm}(BM).  These papers evaluated the leading radiative
corrections to gaugino masses in a class of supersymmetric extentions
of the standard model.  Since then the world-view has changed
considerably, because the top and Higgs are proving to be heavier than
envisaged in those days, and because LEP constraints on new particles
can be brought to bear.  Furthermore the understanding of SUSY and ew
symmetry breaking has advanced enough that much of the
model-dependence of early work can be avoided.  In this note we extend
the BGM/BM analysis, eliminating recourse to a specific model of the
symmetry breaking.  In particular, we avoid their assumptions that
$A=3$ and $\mu = \mtilde$, where $\mtilde$ is the soft SUSY-breaking
mass contribution common to all scalars.  We generalize their results to
arbitrary $tan \beta$ (the ratio of vev's of the two Higgs doublets
which are responsible for electroweak symmetry breaking in
supersymmetric models: $tan \beta \equiv \frac{v_u}{v_d}$).
We also include radiative corrections to the chargino and
neutralino mass matrices which have previously been neglected and which
prove to be important in some regions of parameter space.

Two types of diagrams give the main radiative contributions:
\begin{enumerate}
\item Top-stop loops contribute to the gluino mass and to bino-w3ino
($\bino~\wino_3$) and $\bino~\bino$ entries in the neutralino mass
matrix.  The one loop contribution\cite{bgm,bm} is proportional to the
top mass times a function of the masses of the stop quark eigenstates
$m_{t1}$ and $m_{t2}$, which vanishes when they are degenerate.  There
can also be important 2-loop contributions coming from the
top-stop loop with an additional Higgs exchange if $A$ or $B$ are
non-zero.

\item One loop diagrams containing a W or Higgs and a wino, bino or
higgsino contribute to the $\wino_3-\wino_3$, $\bino-\bino$ and
$\wino_+ - \wino_+$ terms in the neutralino and chargino mass
matrices.  This contribution is proportional to $\mu$, the
SUSY invariant coupling between the two Higgs superfields in the
superpotential, times a function of tree level chargino and neutralino
masses.
\end{enumerate}

\section{Experimental Constraints on Parameters}
\label{charged}
\hspace*{2em}

Since the masses of the charginos and squarks are constrained to be
above about 45 GeV from their non-observation at LEP\footnote{C.f. ref.
\cite{coignet}.  We will also consider below the implications of
lowering the $m_{stop}$ bound to 15 GeV as discussed in ref.
\cite{fmyy}.  We avoid the use of CDF constraints on squark masses at
this stage, since they depend on model dependent properties of other
SUSY particles than the squarks\cite{cdf:gluinolim2}.} the first step
of our analysis is to express these masses in terms of the
parameters $\mu, ~\mtilde,~A,~B$, and $tan \beta$ of the theory, in
order to determine which regions of parameter space are allowed in
this scenario.  In order to make our
analysis independent of the details of the mechanism of ew symmetry
breaking, we do not constrain the parameter space to guarantee the
conditions for radiative electroweak symmetry breaking
mechanism\cite{a-gpw,ir}.

\subsection{Chargino Masses at Tree Level}
\hspace*{2em}

\begin{figure}
\epsfxsize=\hsize
\epsffile{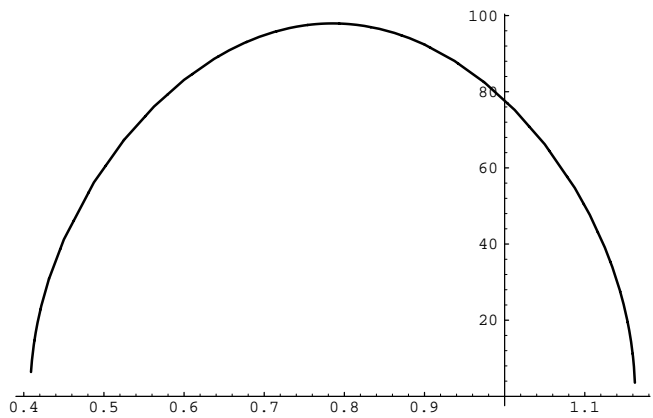}
\caption{Maximum allowed value of $\mu$ in GeV as a function of
$\beta$ from the chargino mass limit. }
\label{fig:mumaxlo}
\end{figure}

We denote the
chargino mass matrix by
\eq
\left( \begin{array}{cc}
0       &       X^T     \\
X       &       0
\end{array} \right)
\en
which acts on the spinor $( \Psi^+ ~~ \Psi^-)$, where $\Psi^{\pm}$ are
two component spinors: $\Psi^+ = (\wino_+ ~~\higgsino_u^+)$ and $\Psi^-
= (\wino_- ~~\higgsino_d^-)$. At tree level, $X$ is the matrix
\eq \left( \begin{array}{cc}

0       &       \sqrt{2}~ m_W~sin \beta \\
\sqrt{2}~ m_W~cos \beta & \mu
\end{array} \right) .
\label{treechargino}
\en
The parameter $\mu$ does not violate supersymmetry.  It enters the
superpotential through the term $\mu \hat{H}_u \hat{H}_d$.  We relate
$X$ to the diagonal matrix ${\cal M }= UXU'$, where $U$ is the
rotation
\eq \left( \begin{array}{cc}
c       &       -s \\
s       &       c
\end{array} \right)
\en
and $U'$ is obtained from $U$ by $c \rightarrow c', s \rightarrow
-s'$.  In terms of the eigenvalues $m_1,~m_2$ of $\cal M$, the tree
level chargino mass matrix, we have
$c \cdot c' = \mu \frac{m_2}{m^2_2 - m^2_1}$ and $s \cdot s' = \mu
\frac{m_1}{m^2_1 - m^2_2}$.  In order that the lighter eigenstate, $m_2$,
is heavier than $m_2^{lim}$, $\mu$ must satisfy
\eq
\mu^2 < \frac{m_W^4 (sin2\beta)^2 - 2 m_W^2 (m_2^{lim})^2 +
(m_2^{lim})^4}{(m_2^{lim})^2} .
\label{mufrchlo}
\en
For $m_2^{lim}=45$ GeV and $\beta = \frac{\pi}{4}$, this gives $\mu
\lsi$ 100 GeV; the limit on $\mu$ is lower for other choices of $\beta$.
Fig. \ref{fig:mumaxlo} shows the upper limit on $\mu$ from the
chargino mass limit, eqn (\ref{mufrchlo}), as a function of $\beta$.

\subsection{Chargino Masses at One-Loop Level}
\hspace*{2em}

Radiative corrections to the chargino mass matrix become significant
if $\mu$ is very large compared to $m_W$.  Then the entries in $X$ are
modified by corrections which can be comparable to the off-diagonal
elements in the tree level matrix.  Taking $\mu$ to be much
larger than any other entry in $X$, the radiatively-corrected mass of
the lighter chargino, $m_2^{rc}$, is essentially equal to the correction to
the $\wino~ \wino$ entry in (\ref{treechargino}).  The main
contribution arises at the one-loop level with charginos, neutralinos,
gauge and Higgs bosons running in the internal lines.  Its exact
expression depends on the detailed mass spectrum of all these
particles.  However since we are interested in the large-$\mu$ limit,
a major simplification occurs.  The higgsinos, $\tilde{h}_u$ and
$\tilde{h}_d$ combine together to form a Dirac SU(2) doublet of mass
$\mu$, while the light eigenvectors of the chargino and neutralino mass
matrices are mainly gauginos.  A similar simplified
pattern occurs also in the scalar Higgs sector.  It is
known\cite{bm:custSU2} that, in the limit where one switches off the
U(1) hypercharge coupling, the neutral Higgs scalar potential exhibits
an SU(2)$_L$xSU(2)$_R$ global invariance.  The ew breaking
breaks this global symmetry.  However if $\mu$ and/or
$\mtilde$ are $ >> m_W$, the corrections to the above global symmetry
are small and we can still classify the spin-0 mass eigenstates into
two approximate SU(2) doublets.  They are obtained from linear
combinations of $H_u$ and $H_d$, suitably weighted by $cos \beta$ and
$sin \beta$ coefficients.  One combination contains the massless
SU(2)xU(1) would-be-Goldstone bosons and one light neutral Higgs
(whose mass is $\approx m_W$).  Its couplings to the external gauginos
are fixed by the Higgs mechanism.  The orthogonal combination contains
the heavy charged and neutral Higgs bosons, whose couplings are fixed
by the orthogonality condition.  If we denote by $M$ the mass of these
latter bosons, $M$ will be of order of the larger of $\mu$ and
$\mtilde$.

\begin{figure}
\epsfxsize=\hsize
\epsffile{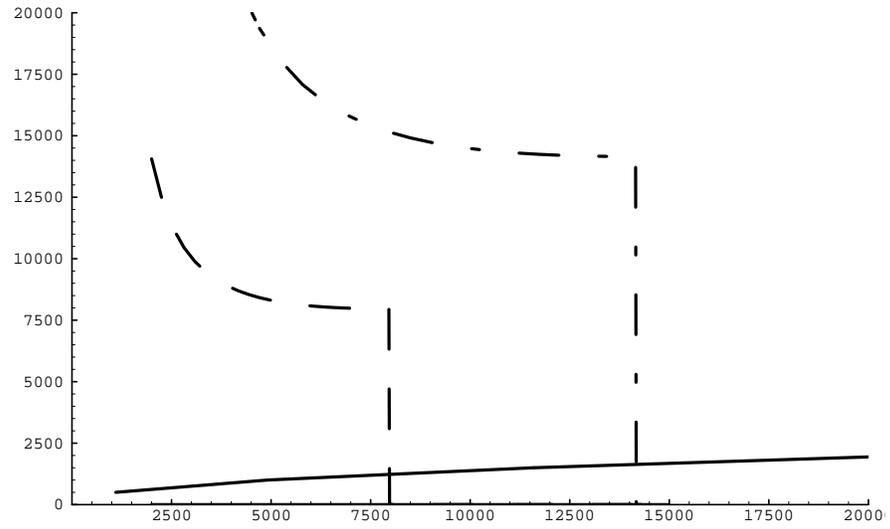}
\caption{Lower limits on $M$ as a function of $\mu$ for
consistency with a chargino mass limit of 45 (dashed) and 80
(dot-dashed) GeV.  Solid curve is the boundary of the region allowed
by the squark mass limit, with the allowed region being above the
curve.  All masses are in GeV in this and other figures.}
\label{fig:mmin}
\end{figure}

Making use of the above mass spectrum, one obtains the following
one-loop contribution to the $\wino \wino$ entry in
(\ref{treechargino}):
\eq
\delta_{\wino \wino} = \frac{\alpha_2}{2 \pi} \frac{\mu m_1 m_2}{m_1^2
- m_2^2} \left( 3 F(m_W,m_1,m_2) + \frac{m_2^2}{m_W^2} F(m_2,m_1,M)
- \frac{m_1^2}{m_W^2} F(m_1,m_2,M) \right),
\label{delta33}
\en
where
\eq
F(x,y,z) =     \frac{z^2}{x^2-z^2} log[\frac{z^2}{x^2}]
-\frac{y^2}{x^2-y^2} log[\frac{y^2}{x^2}].
\label{f}
\en
When $\mu$ and $M$ are much larger than $m_2 \lsi m_W$,
this gives approximately
\eq
m_2^{rc} = \frac{\alpha_2}{ \pi}cos \beta sin \beta \frac{\mu M^2}
{(\mu^2-M^2)} log [ \frac{\mu^2}{M^2} ].
\label{m2rcapprox}
\en
This expression shows that the present experimental
bound on the lighter chargino can be accomodated for sufficiently
large $\mu$ and $M$.  As noted above, when $\mtilde >> \mu$, we expect
$M \sim \mtilde$, while if $\mtilde << \mu$ we expect $M \sim \mu$.
Simply using eqn (\ref{m2rcapprox}) with $M = max[\mtilde,\mu]$ gives
the boundaries of the allowed regions of $\mu$ (on the horizontal
axis) and $\mtilde$ (on the vertical axis) shown in
Fig. \ref{fig:mmin}, for $\beta = \frac{\pi}{4}$.  Once $\mu \gsi
\frac{\pi m_2^{lim} }{ \alpha_2 cos \beta sin \beta} $, the condition
(\ref{m2rcapprox}) is satisfied for any $\mtilde$, accounting for the
vertical segments.  For the present value of $m_2^{lim}=45$ GeV we
have the dashed curve, while the dot-dashed curve gives the boundary
of the allowed region for $m_2^{lim}=80$ GeV.  The allowed region is
above and to the right of these curves, but remember that the sharp
corners and vertical lines are artifacts of the simplistic relation $M
= max[\mtilde,\mu]$.  Given a particular model, one can find the
smooth curve which this approximates.  For other values of $\beta$,
the boundary shown in the figure should be multiplied by $[2 cos\beta
sin\beta]^{-1}$.  For $\mu \lsi 2$ TeV one sees that the $M$ required
becomes very large: for $\mu=2$ (1.5, 1) TeV, $M$ must be larger than
16 (27, 94) TeV, respectively.  Thus in this tree-level-massless
gaugino scenario, unless one is willing to consider very large values
of $M$ there is effectively a gap in allowed $\mu$'s between $\sim
100$ GeV and a few TeV.\footnote{If the tree-level mass of charginos
is of O(50-90) GeV, as studied in the recent preprint
\cite{strassler}, then the radiative corrections considered here can
be large enough to be experimentally significant for smaller values of
$\mu$.  GF thanks M. Strassler for a discussion of this work.} If no
chargino is found at LEPII and $m_2^{lim}$ is increased to above
$m_W$, then the entire low-$\mu$ region will be removed and the
allowed range of $\mu$ and $M$ is just the area above the dot-dashed
curve in Fig. \ref{fig:mmin}.

\subsection{Stop Mass}
\hspace*{2em}

The top squark $mass^2$ matrix for the effective low-energy theory is\\
approximately\footnote{The $mass^2$ matrix (\ref{mstop}) takes the
low-energy soft-susy-breaking contributions to the
$\tilde{t}\tilde{t}$ and ${\tilde{t}}^c{\tilde{t}}^c$ entries equal to
a single parameter $\mtilde^2$.  Modification of this simplest
assumption, e.g., due to RG running, is discussed below.}:
\eq
\left( \begin{array}{cc}
m^2_t + \mtilde^2 + m_Z^2 cos (2 \beta) (\frac{1}{2} - \frac{2}{3} sin^2
\theta_w)       &       A_{\rm eff} m_t \mtilde + \mu m_t cot \beta \\
A_{\rm eff} m_t \mtilde + \mu m_t cot \beta   & m^2_t + \mtilde^2 +
m_Z^2 cos (2 \beta) ( \frac{2}{3} sin^2 \theta_w)
\end{array} \right),
\label{mstop}
\en
where
\eq
A_{\rm eff} \equiv A + \delta A + B \frac{k \mu^2}{M^2}
\label{Aeff}
\en
and $\delta A$ is the radiative correction to $A$ from the gluino-top
loop, in which the gluino mass insertion is a one-loop diagram.  When
$A=B=0$ at tree level this correction can be relevant for some regions
of parameters.  From formulae given in ref. \cite{a-gpw}, its magnitude
is $\delta A = \frac{4 \alpha_s}{\pi} \frac{m_{\gluino}}{\mtilde}
log(\frac{M_{initial}}{m_W})$.  If $B \ne 0$, a $\tilde{t} \tilde{t}^c
$ mixing can also arise from the vertices $h_t \mu \tilde{t}
\tilde{t}^c H_d$ and $B \mtilde \mu H_u H_d$ connected by an $H_d$
propagator.  Evaluating this propagator at zero momentum
and taking the vev of $H_u$ produces a contribution to $A_{\rm eff}$
proportional to $B$. Its sign and precise magnitude depend on the
details of the Higgs mass spectrum, so we parameterize it in terms of
$k$, a constant which is presumably of order one, and a generic scalar
Higgs mass, $M$.

When dimension-3 SUSY-breaking operators are absent from the low energy
theory, $A=0$; more commonly it has been taken to be of order 1, e.g.,
3 in BM\cite{bm}.  $A_{\rm eff}$ cannot be made too large or
the scalar quarks or leptons will get a vev and color SU(3) or electromagnetism
will be broken.  Typically this leads to an upper bound on the modulus
of $A$ close to 3\cite{fjr,a-gpw,il,dgg}.  We will see below that
consistency with the experimental lower bound on the lighter stop
mass generally requires $A_{\rm eff}$ to be even smaller than this.

The diagonal terms in the stop $mass^2$ matrix determine the average
squark $mass^2$.  Splitting between the physical stop mass eigenstates
is mainly controlled by the off-diagonal terms, as long as the
diagonal terms are not too different.  Thus an experimental lower
limit on the stop mass, $m_{stop}^{lim}$, implies an upper limit on
$A_{\rm eff} \mtilde + \mu cot \beta$ for a given average stop
mass-squared.  Dropping the small $m_Z^2$ corrections in
(\ref{mstop}) to make the point clear, this is
\eq
A_{\rm eff} \mtilde +  \mu cot\beta \lsi \frac{\mtilde^2 + m_t^2 -
(m_{stop}^{lim})^2}{m_t}.
\label{mucotbetamax}
\en
{}From this expression one sees that the limit on $\mu$ from
the stop mass constraint is essentially independent of $m_{stop}^{lim}$ and
$A_{\rm eff}$ when $\mtilde$ is large.  The upper limit on $\mu$ for a
given $\mtilde$ is shown in the large $\mu$ region as the solid line
Fig. \ref{fig:mmin}; the allowed region of $\mtilde$ for a given $\mu$
is above the line.  One sees that in the large $\mu$ region if the chargino
constraint is satisfied, the squark constraint will usually be also.

In the small $\mu$ region when $tan \beta \ge 1$, the constraint
from the stop mass limit is less stringent than from the chargino
limit, except for very small $\mtilde$.  Since the CDF
limits on squark masses must be reexamined when the gluino becomes as
light as we will be considering, we use $m_{stop}^{lim}=45$ GeV to be
conservative.  However even if the strongest CDF limit of
$m_{stop}^{lim} = 126$ GeV were applicable, we found that it would
make an insignificant difference in these limits except for $A_{\rm
eff} \ne 0$ and small $\mtilde$.  We have checked that modifying the
$\mtilde^2$ terms in the diagonal elements of (\ref{mstop}) as would
arise from different renormalization group running of the $\tilde{t}$
and ${\tilde{t}}^c$ masses in the RG-induced ew symmetry breaking
scenario, does not significantly affect these conclusions, again
because the chargino mass provides the more stringent constraints on
parameters.

\begin{figure}
\epsfxsize=\hsize
\epsffile{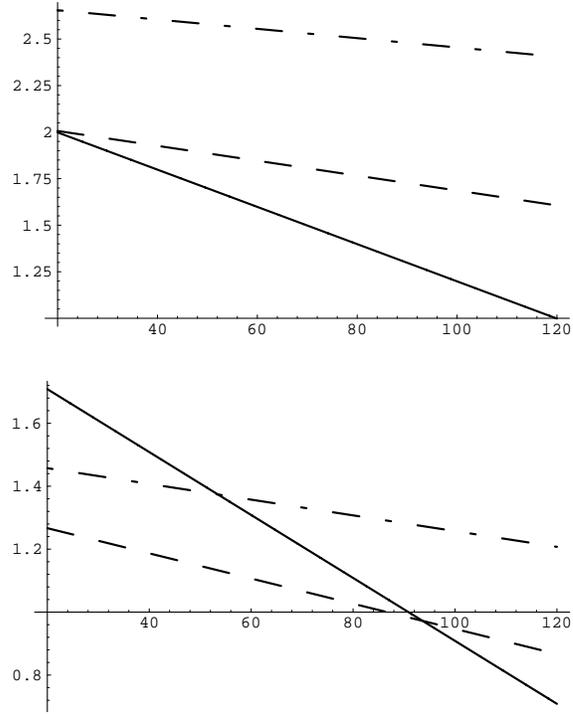}
\caption{Maximum allowed value of $A_{\rm eff}$ in the low $\mu$ region
as a function of $\mu cot \beta$ consistent with $m_{stop}^{lim}=45$
GeV, for $\mtilde = 100$ (solid), 250 (dashed), and 400 GeV
(dot-dashed). Upper plot uses the stop $mass^2$ matrix with equal
susy-breaking diagonal terms; lower plot uses modified diagonal
entries as discussed in the text.}
\label{fig:amax}
\end{figure}

Thus for most of the interesting parameter space in the small as well
as large $\mu$ region, consistency with the LEP chargino and squark
mass limits is guaranteed simply by satisfying eqn (\ref{mufrchlo})
from the chargino limit, independent of the stop mass limit,
$\mtilde$, and $A_{\rm eff}$ (as long as it is not too large).  Note
however that for larger $A_{\rm eff}$ the stop mass limit becomes
dominant and in fact requires that $A_{\rm eff}$ be less than some
maximum value for given $\mtilde$ and stop mass
limit. Fig. \ref{fig:amax} shows this, for $\mtilde = 100$ (solid),
250 (dashed), and 400 GeV (dot-dashed).  The upper plot uses the stop
$mass^2$ matrix (\ref{mstop}), while for the lower plot the $\mtilde^2$ in
the (1,1) and (2,2) element of (\ref{mstop}) has been modified to
$\frac{2}{3}\mtilde^2$ and $\frac{1}{3}\mtilde^2$, respectively.  This
simulates (see Table 1 of ref. \cite{a-gpw}) the case that these terms
are equal at the susy-breaking scale but experience RG running
which also causes the ew gauge symmetry to break.

To summarize, requiring the lightest SUSY charged particles to be
heavier than the experimental lower bounds leads to two distinct
allowed regions for $\mu$ and associated regions for $\mtilde$ --
namely $\mu \lsi 100$ GeV, or $\mu \gsi$ several TeV.  Now let us find
the gluino and lightest  neutralino ($\chi^0_1$) masses for the
allowed parameter regions.
\newpage

\section{The Gluino}
\label{sec:gluino}
\hspace*{2em}

The top-stop loop produces the only important 1-loop correction to the
gluino mass\cite{bgm,bm}:
\eq
\delta^{(1)}_{\gluino} = \frac{\alpha_s m_t}{4 \pi} sin (2 \theta_t)
F(m_t,m_{t1},m_{t2}),
\label{delta1gl}
\en
where the function $F$ is the same as in eqn (\ref{f}) and $\theta_t$
is the rotation which diagonalizes the stop mass matrix\footnote{We
thank D. Pierce for pointing out that we had omitted writing this
factor in the original version of the manuscript; it has been included
in the numerical analysis.}.  Using (\ref{mstop}) for the stop
$mass^2$ matrix,
\eq
sin^2 (2 \theta_t) = \frac{[(A_{\rm eff} \mtilde + \mu cot
\beta) m_t]^2}{[(A_{\rm eff} \mtilde + \mu cot
\beta) m_t]^2 + \frac{1}{4}[m_Z^2 cos(2 \beta)(\frac{1}{2} - \frac{4}{3}
sin^2 \theta_W)]^2}.
\label{sin2thetat}
\en
In this case, $sin (2 \theta_t) \approx 1$ for most of parameter
space.  We will consider below the case that (\ref{mstop}) is modified
such that the soft-susy breaking contributions to the diagonal
elements of the stop $mass^2$ matrix are not equal.  Note that
$F(x,y,z)$ is odd under $ y \leftrightarrow z$ so that
$\delta^{(1)}_{\gluino} $ can be seen to vanish linearly with the
fractional splitting between the stop mass eigenstates.  Having
$A_{\rm eff}$ non-zero or having a large value of $\mu cot \beta$
contributes to a larger gluino mass because each of these increases
the mass splitting between stop mass eigenstates (see eqn (\ref{mstop})).

The top-stop contribution to gaugino masses can have a 2-loop
divergent piece coming from Higgs exchange between top and stop, if
the dimension-3 SUSY-breaking scalar trilinear coupling $A \mtilde
\tsquark \tsquark^c H_2$ is non-vanishing.  All divergent 2-loop
diagrams have been calculated recently in refs. \cite{yamada,mvtwoloops}
and we use their result here.\footnote{The newer results differ by an
overall factor of 3 from that given in \cite{bm}. Note that \cite{bm}
corrects a factor-of-2 error in the one-loop contribution given in
\cite{bgm}.}  Denoting by $M_{initial}$ the renormalization scale at
which the counterterm exactly cancels the contribution of this
divergent graph, so that gauginos are massless, the RG contribution to
the low energy gluino mass is:
\eq
\delta^{(2)}_{\gluino} = \frac{\alpha_3 \alpha_2 (1 + cot^2 \beta)
}{4 \pi^2}  A \mtilde
\left(\frac{m_t}{m_W}\right)^2log(\frac{M_{initial}^2}{m_W^2}).
\label{delta2gl}
\en

In addition, if $B$ is non-zero, there is a finite 2-loop contribution
to the gluino mass which can be important for some portions of parameter
space.  It corresponds to the same diagram as the one just considered,
but with the pointlike vertex $A \mtilde \tilde{t} \tilde{t}^c H_u$
replaced by the vertices $h_t \mu \tilde{t} \tilde{t}^c H_d$ and $B
\mtilde \mu H_u H_d$ connected by an $H_d$ propagator.  For $\mu$ and
$\mtilde>> m_W$ and $m_t$, this correction is approximately
\eq
\delta^{(3)}_{\gluino} \approx \frac{\alpha_3 \alpha_{em}}{4 \pi^2
sin^2\theta_W}  \frac{B \mu^2 \mtilde}{\bar{M}^2}
\left(\frac{m_t}{m_W}\right)^2,
\label{delta3gl}
\en
where $\bar{M}$ denotes the highest mass in the loop.  Taking
$\bar{M} = max[\mu,\mtilde]$, this contribution is maximized for $\mu
= \mtilde$, for which it is $4 \times 10^{-4} \mu B$.  Thus it is only
relevant if $A=0$ and $\mu  \sim \mtilde$, with $ \mu B$ of order
several TeV or larger.

\begin{figure}
\epsfxsize=\hsize
\epsffile{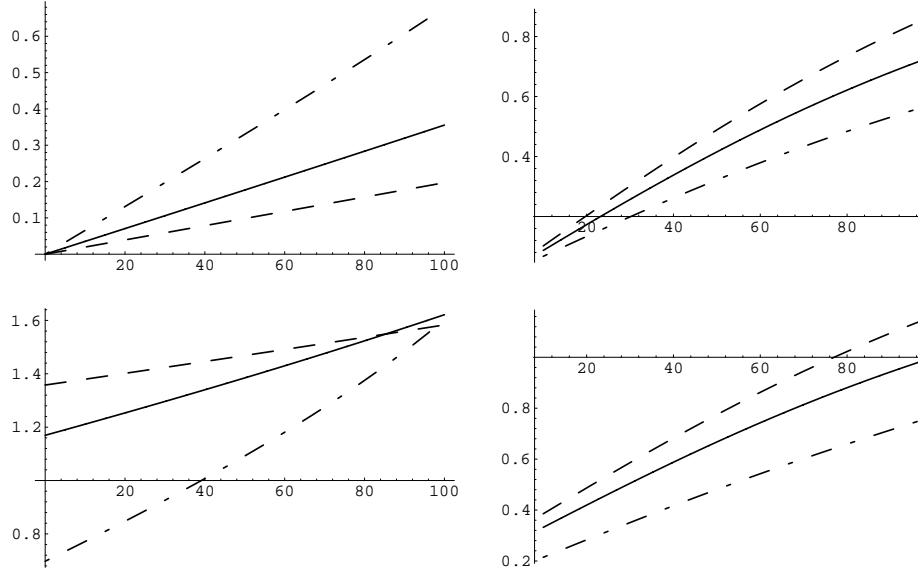}
\caption{Gluino (left column) and $\chi^0_1$ (right column) masses in
GeV as a function of $\mu$ in GeV, for $A_{\rm eff}=0$ (upper row) and
$A_{\rm eff}=1$ (lower row), with $\mtilde = 100$ (dot-dashed), 250(solid)
and 400 GeV(dashed); $\beta = \frac{\pi}{4}$.}
\label{fig:mmu}
\end{figure}

\begin{figure}
\epsfxsize=\hsize
\epsffile{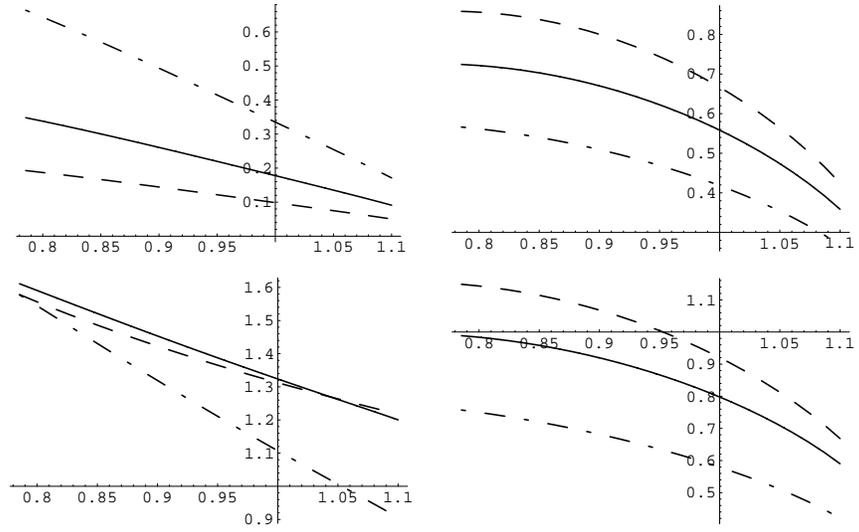}
\caption{Gluino (left column) and $\chi^0_1$ (right column) masses in
GeV as a function of $\beta$, with $\mu$ taken at its maximum value
consistent with the charginos and stops being heavier than 45 GeV.
For $A_{\rm eff}=0$ (upper row) and $A_{\rm eff}=1$ (lower row), with $\mtilde
= 100$ (dot-dashed), 250(solid) and 400 GeV(dashed).}
\label{fig:mbeta}
\end{figure}

\begin{figure}
\epsfxsize=\hsize
\epsffile{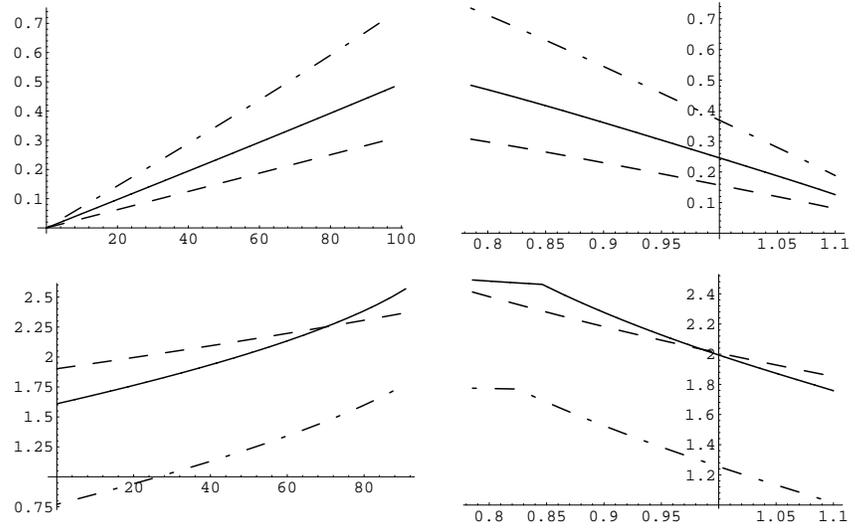}
\caption{Gluino masses when diagonal SUSY breaking stop $mass^2$ terms
are not taken equal, as described in the text.  Left
(right) column of plots are as in Fig. 4 (5).}
\label{fig:mgl_rg}
\end{figure}

The first column of Fig. \ref{fig:mmu} shows the one-loop contribution
to the gluino mass, $\delta^{(1)}_{\gluino}$, for $A_{\rm eff}=0$ and
$A_{\rm eff}=1$, as a function of $\mu$ in the low $\mu$ region, for
$\beta = \frac{\pi}{4}$ and several choices for $\mtilde$.  Also in
the low $\mu$ region, the first column of Fig. \ref{fig:mbeta} shows
the gluino mass as a function of $\beta$ at the maximum value of $\mu$
which is consistent with whichever is the stronger of the stop mass or
chargino mass constraints (in fact, almost always the latter), for
$A_{\rm eff}=0$ and 1.  These results are computed with the stop
$mass^2$ matrix (\ref{mstop}).  If there are significant differences
in the low-energy soft-susy-breaking diagonal terms in the stop
$mass^2$ matrix, the gluino mass predictions are modified somewhat.
To illustrate the possible extent of this effect, consider the
scenario of radiative ew symmetry breaking.  In that case, the
$\mtilde^2$ in the $1,1$ and $2,2$ elements of (\ref{mstop}) is
multiplied by $\sim 2/3$ and $\sim 1/3$ respectively, taking the
values of the corrections chosen in the previous section as an example.
For $\mtilde \gsi m_t$ the change is quantitatively although not
qualitatively important.  We give the 1-loop contributions to the
gluino mass predictions for this case in Fig. \ref{fig:mgl_rg}.

\begin{figure}
\epsfxsize=\hsize
\epsffile{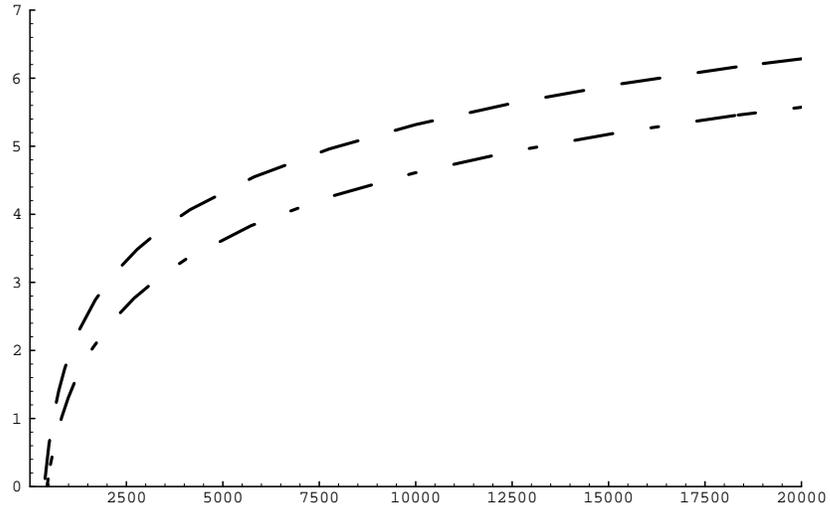}
\caption{Maximum possible one-loop contribution to the gluino mass in
the large-$\mu$ region when $A_{\rm eff}=0$ and $\beta= \frac{\pi}{4}$,
plotted versus $\mu$, corresponding to taking the lighter stop mass to
be 45 GeV (dashed) and 126 GeV (dot-dashed) respectively. }
\label{fig:mglmaxhiA0}
\end{figure}

For $\mtilde \gsi m_t$, the 1-loop contribution to the gluino mass
decreases as $\mtilde$ is increased with $A$ and $\mu$ held fixed.
This is because increasing $\mtilde$ decreases the fractional splitting
between the stop mass eigenstates, $\sim \frac{(A_{\rm eff} \mtilde + \mu cot
\beta)m_t}{(\mtilde^2 + m_t^2)}$.  Thus for $A=B=0$ the gluino mass is
negligible in the large $\mu, ~ \mtilde$ region, unless $\mu m_t \sim
{\mtilde}^2$.  The maximum value of the gluino mass in this latter case
occurs when the the lighter stop is as light as is allowed
experimentally while the heavier stop is very massive, thus maximizing
the fractional splitting between eigenstates.  Figure \ref{fig:mglmaxhiA0}
shows the maximum gluino mass under these circumstances, with
$m_{stop}$ greater than 45 (dashed) and 126 (dot-dashed) GeV.
The  relationship required to implement this, $\mu \approx
\mtilde^2/m_t >> \mtilde$, is unconventional.

\begin{figure}
\epsfxsize=\hsize
\epsffile{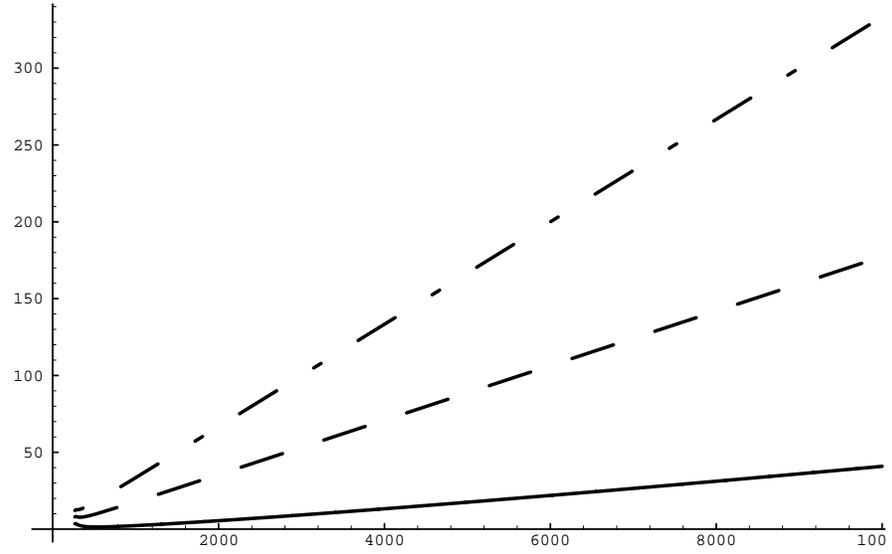}
\caption{Gluino mass from the two-loop divergent diagram when $A=1$, $B=0$
and $\beta= \frac{\pi}{4}$ versus $\mtilde$ up to 10 TeV, for
$M_{initial} = \mtilde$ (solid), $M_{initial} = 10^{11}$ GeV (dashed),
and $M_{initial} = M_{Pl}$ (dot-dashed).}
\label{fig:mgl2loop}
\end{figure}

If $\mtilde$ is large and $A \ne 0$ the divergent 2-loop contribution can be
important.  Fig. \ref{fig:mgl2loop} shows $\delta^{(2)}_{\gluino}$ for
$A=1$, $B=0$ and $\beta = \frac{\pi}{4}$.  The solid curve corresponds to
taking $M_{initial} \rightarrow \mtilde$, giving an estimate of the
minimal importance of this correction, while the dashed and dot-dashed
curves show the result for $M_{initial} = 10^{11}$ GeV and
$M_{initial} = M_{pl}$.  Evidently, for large $M_{initial}$ this is
a large effect.  It would be interesting to determine the value of
this two-loop contribution imposing as well the constraints of the
radiative electroweak breaking scenario.

To summarize, for the 3 cases we are treating,
\begin{enumerate}

\item $A=B=0$:  In the low $\mu$ region the gluino mass decreases with
increasing $\mtilde$ from $\lsi 700$ MeV for $\mtilde = 100$ GeV to
$\lsi 200$ MeV for $\mtilde = 400$ GeV (Fig. \ref{fig:mbeta}, upper
left, and Fig. \ref{fig:mgl_rg}, upper right).  In the large $\mu$
region the gluino mass is negligible unless $\mu \sim \mtilde^2/m_t$,
in which case the maximum gluino mass is $\sim 6 $ GeV for $\mu \lsi
20$ TeV (Fig. \ref{fig:mglmaxhiA0}).

\item $A=0,~ B \ne 0$:  When $\mu << \mtilde$, this case is equivalent
to the previous case with $A=B=0$.  For $\mu \sim \mtilde$ in the low
$\mu$ region, the $kB \frac{\mu^2}{M^2}$ contribution to $A_{\rm eff}$
can produce $A_{\rm eff} \sim 1$ so that guino masses can be of order
a few GeV (see Fig. \ref{fig:mbeta}, lower left plot, and Fig.
\ref{fig:mgl_rg}, lower right plot).  For the large
$\mu \sim \mtilde$ region the two loop diagram proportional to $B$
makes a contribution (eqn \ref{delta3gl}) $\sim 4 \times 10^{-4} \mu
B$.

\item $A \ne 0$:  In the low $\mu$ region this gives gluino masses of
order a few GeV as discussed in the item above.  However in the large
$\mu$ region the gluino mass can be very large due to the 2-loop
divergent diagram: e.g., for $A=1$ and $M_{initial} \gsi 10^{11} $
GeV, the gluino mass is consistent with the present CDF missing energy
bound\cite{cdf:gluinolim2} as long as $\mu \gsi 8$ TeV (see
Fig. \ref{fig:mgl2loop}).
\end{enumerate}

\section{The Lightest Neutralino}
\hspace*{2em}

The tree level neutralino mass matrix, in the basis
$(\bino,\wino_3,\higgsino_1,\higgsino_2)$, is:
\eq
\left( \begin{array}{cccc}
0       & 0     &       -M_Z cos \beta sin \theta_w     & M_Z sin \beta sin
\theta_w \\
0       & 0     &       M_Z cos \beta cos \theta_w      & -M_Z sin \beta cos
\theta_w \\
-M_Z cos \beta sin \theta_w     & M_Z cos \beta cos \theta_w    & 0     & -\mu
\\
M_Z sin \beta sin \theta_w      & -M_Z sin \beta cos \theta_w   & -\mu  & 0
\end{array} \right).
\label{neutralino}
\en
Radiative corrections remove the zeros in this matrix.  Let us first
consider the radiative contributions to the neutral gaugino 2x2
sub-matrix in the upper left-hand corner.

The $\bino-\wino_3$ off-diagonal entries receive one- and two-loop
contributions entirely analogous to those that we computed for the
gluino mass:
\eq
\delta_{\bino \wino} = \frac{\sqrt{\alpha_1 \alpha_2}}{\alpha_3}
m_{\gluino},
\label{deltab3}
\en
where $m_{\gluino} = \delta^{(1)}+\delta^{(2)}+\delta^{(3)}$, given in eqns
(\ref{delta1gl}),(\ref{delta2gl}) and (\ref{delta3gl}).  As for the
diagonal entries,
the contribution to $\wino_3-\wino_3$ is readily related to
$\delta_{\wino \wino}$ in the chargino sector, eqn (\ref{delta33}), in
the approximation of large $\mu$ that we discussed there.  Finally,
the $\bino \bino$ entry receives two types of radiative
contributions.  The first comes from one- and two-loop corrections
with top and stop running in the loops, yielding a contribution
proportional to $m_{\gluino}$ analogous to the expression in eqn
(\ref{deltab3}).  The other type of correction is from higgsino-higgs
loops\footnote{In Feynman 'tHooft gauge, where diagrams with gauge bosons in
the loop vanish.}. It is the same as for $\wino
\wino$, replacing $\alpha_2$ by $\alpha_1$.  All together we obtain:
\eq
\delta_{\bino \bino} = \frac{2 \alpha_1}{3 \alpha_3} m_{\gluino} +
\frac{\alpha_1}{\alpha_2} \delta_{\wino \wino}.
\label{deltabb}
\en

We do not compute the radiative corrections to the higgsino submatrix
in detail, since they depend on the model of ew symmetry breaking.
They would not be present if there were a Peccei-Quinn symmetry, so
that they must be proportional to $\mu$ and/or the vev's of $H_u$ and
$H_d$.  We find that these radiative corrections cannot be larger than
the $O(\mu)$ entries which are present in eqn (\ref{neutralino}).  We
checked that the masses of the lightest two neutralinos change only
slightly when such terms are included.

\begin{figure}
\epsfxsize=\hsize
\epsffile{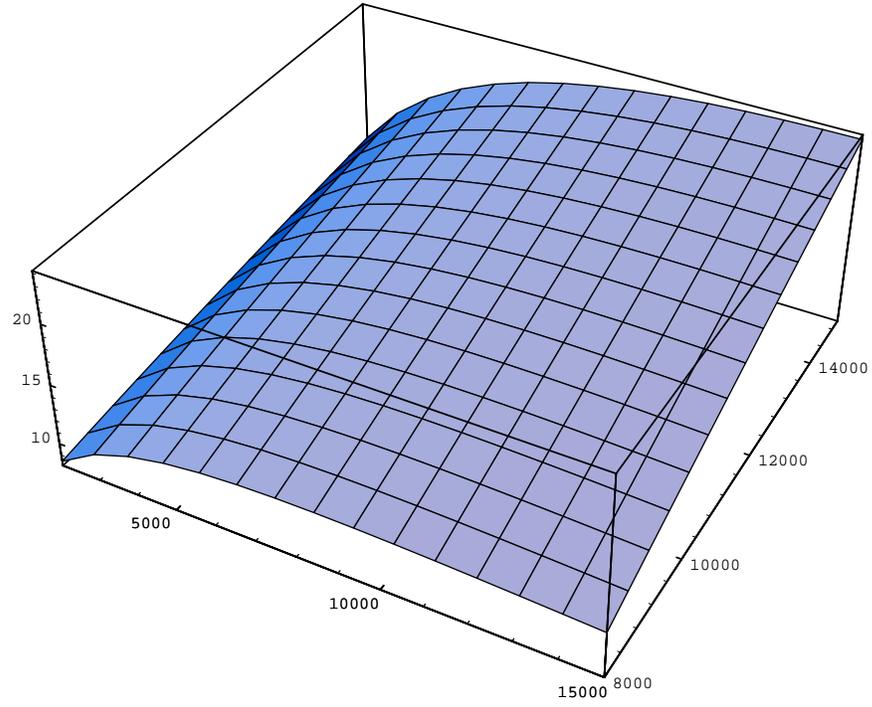}
\caption{Mass of the lightest neutralino versus $\mu$ and $M$ with
$\mu$ running from 2 TeV to 15 TeV and $M$ from 8 TeV to 15 TeV,
taking $M_{initial}=\mtilde$.}
\label{fig:mlsphi}
\end{figure}

We find the eigenvalues and eigenvectors of the radiatively-corrected
neutralino mass matrix numerically, for a variety of values of
parameters.  The mass of the lightest neutralino, $\chi^0_1$, in the
small $\mu$ region is shown in the second column of Figs. \ref{fig:mmu} and
\ref{fig:mbeta} as a function of $\mu$ and $\beta$, respectively.  In
the small $\mu$ region, $m(\chi^0_1)$ is rather insensitive to
$\mtilde$ and $A_{\rm eff}$, but is sensitive to $\mu$.  Typical
values for $m(\chi^0_1)$ in the small-$\mu$ case are
$O(\frac{1}{10}-1)$ GeV.  Since the top-stop loop is responsible for
only a fraction of the neutralino mass, $m(\chi^0_1)$ is insensitive
to possible differences in the susy-breaking diagonal stop
$mass^2$ and the analog of Fig. \ref{fig:mgl_rg} is not needed.  The
dependence of $m(\chi^0_1)$ on $\mu$ and $\mtilde$ in the large $\mu$
region is shown in Fig. \ref{fig:mlsphi}. There one sees that
$m(\chi^0_1) \gsi 10 $ GeV for the large-$\mu$ case.  The sensitivity
to $A_{\rm eff}$ is too small to be seen in this scale figure.  The
second-lightest neutralino, $\chi^0_2$, typically has a mass of about
50 GeV.

The gluino mass is more strongly dependent on $A_{\rm eff}$ than are
the neutralino masses just because the gluino gets its mass entirely
from quark squark loops which are sensitive to $A_{\rm eff}$ and
proportional to the gauge coupling constant appropriate the the
gaugino in question.  On the other hand, the bino-wino submatrix
of the full neutralino mass matrix, whose eigenvalues are dominantly
important in determining the lightest neutralino mass, is approximately
\eq
\left( \begin{array}{cc}
\frac{1}{15}m_{\gluino} + \frac{1}{4} m_2^{rc}      &
\frac{1}{5} m_{\gluino}     \\
\frac{1}{5} m_{\gluino}      &       m_2^{rc}
\end{array} \right) ,
\label{wbsubmatrix}
\en
when the known gauge couplings are inserted into eqns (\ref{deltab3})
and (\ref{deltabb}).  Evidently, the eigenvalues of (\ref{wbsubmatrix})
are insensitive to the top-stop loops unless the radiatively generated
gluino mass is $\gsi 4 m_2^{rc}$.  Since we are only considering
parameter ranges such that $m_2^{rc} > 45$ GeV, the lightest
neutralino mass is generically insensitive to $A_{\rm eff}$ unless
$m_{\gluino}\gsi 200$ GeV.  One can also see from (\ref{wbsubmatrix})
how restricting the parameter space further by improving the chargino
mass limits would in general simply scale up the predictions for the
masses of the lightest neutralinos in proportion to the chargino mass
limit.

For parameters such that $m(\chi^0_1) \lsi 2$ GeV the composition of
$\chi^0_1$ and $\chi^0_2$ are insensitive to parameters and
\eq
|\chi^0_1 > \approx 0.88 |\bino > + 0.47 | \wino_3 >.
\en
This is very close to the U(1)xSU(2) composition of the photon, so in the
small $\mu$ region, the lightest neutralino is essentially a photino.
The $\bino$ component becomes more dominant with increasing
$m(\chi^0_1)$, reaching about $0.99$ for the large $\mu$ scenario.  In
all cases, however, the higgsino components have amplitudes less than
$1\%$ for both $\chi^0_1$ and $\chi^0_2$.  This explains the
insensitivity of the masses of the two lightest neutralinos to the
model-dependent radiative corrections to the higgsino mass submatrix
noted in the previous paragraph.  Since the $Z^0$ only decays to
neutralinos through their higgsino components, the relative
probability of a $Z^0$ decaying to a pair of neutralinos, compared to
decaying to a given neutrino-antineutrino pair, is $\lsi 10^{-8}$.
Thus the impressive experimental constraint from LEP on the number of
extra neutrinos is insufficient to limit the existance of these
neutralinos.

\section{Phenomenology and Cosmology}
\hspace*{2em}

Now we briefly turn to the phenomenological viability of
the scenario we have investigated.  While our analysis above was
general enough to include arbitrary $A$, it is particularly
interesting to consider $A=0$.  This
is because in hidden sector dynamical SUSY breaking without gauge
singlets, all dimension-3 SUSY-breaking operators in the low
energy theory, including a gaugino mass term and the trilinear
squark-squark-Higgs coupling whose coefficient is defined to be $A
\mtilde$, are suppressed by a factor $\frac{\mtilde}{M_{pl}}$ and thus
are expected to be very small\footnote{See ref. \cite{bkn} for a more
detailed discussion of the argument.}.  As long as $A_{\rm eff}$ is
small, the lightest neutralino is generically heavier than the gluino.
For instance for $tan \beta = 1$ and $\mu = 100$ GeV, the lightest
neutralino mass falls in the range $0.5-0.8$ GeV, while the gluino
mass is found to be less than 0.3 GeV (see Fig. \ref{fig:mmu}).  In
the large $\mu$ region the
lightest neutralino mass is greater than $10$ GeV.  Throughout the
large $\mu$ region the upper limit on the gluino mass consistent with
the experimental lower limit on the stop and chargino masses is less than the
lightest neutralino mass\footnote{Unless $\mu$ and $\mtilde$ are {\it very}
large so that $B$ terms can dominate.}.

The phenomenology of hadrons containing light gluinos is discussed in ref.
\cite{f:95} and references cited therein.  Some essential conclusions
are the following:
\begin{enumerate}
\item The theoretical lower limit on the gluino mass coming
from requiring that the $\eta'$ be a pseudogoldstone boson is
$m_{\gluino}\sim 10 \frac{<q \bar q>}{<\lambda \lambda>}
m_s$\cite{f:95}.  The gluino condensate is very uncertain but is
expected to be larger than the quark condensate.  Conceivably the
ratio is large enough to cancel the factor of 10, leading to a lower
bound on the gluino mass of order one to several hundred MeV.  This is
just the range found above in the low $\mu$ region for $A_{\rm
eff}=0$, so that improvements in the determination of the $\eta'$ mass
as a function of the mass of a light gluino will allow part of the
parameter space to be excluded.  For $A=B=0$, one can already exclude
$\mtilde \gsi 300$ GeV when $\mu \lsi 100$ GeV.
\item The non-observation\cite{cusb} of any peak in the photon
spectrum in radiative $\Upsilon$ decay excludes gluinos in the mass
range $\sim 1.5 - 3.5$ GeV, for any lifetime.  This excludes small
regions of parameter space in the large $\mu$ region.
\item Light gluinos would be mainly found in the flavor-singlet hadron
$R^0$, a gluon-gluino bound state, or the flavor-singlet baryon $S^0$
composed of $uds \gluino$.  The mass of the $R^0$ can be
estimated\cite{f:95} from the lattice calculation of the mass of the
$0^{++}$ glueball to be $1440 \pm 375$ MeV for a massless gluino.
$R^0$'s with mass $\lsi 2.2$ GeV are experimentally allowed, except for
lifetimes in the $ \sim 2 \times 10^{-6}-10^{-8}$ sec
range\cite{bernstein} or shorter than $\sim 5 \times 10^{-11}$ sec,
where beam dump experiments are useful\cite{f:95} if $\mtilde$ is not
too large.  In the small $\mu$
scenario the lightest neutralino is typically heavier than the gluino,
and the $R^0$ decay rate is suppressed compared to the conventional
phenomenological treatment in which the lightest neutralino is assumed
to be essentially massless.  Suitable methods to estimate the $R^0$
lifetime must be developed to see if the present experimental limits
constrain this scenario.
\item  Long lived or absolutely stable $R^0$ and $S^0$ are not
obviously excluded.
They would not bind to nuclei, so would not be found in
searches for exotic isotopes\cite{f:95}.  In fact, they could
help provide the dark matter of the universe and might account for
anomalous production of muon events by cosmic rays coming from Cygnus
X-3\cite{f:95}.  Stable or very long-lived $R^0$ and $S^0$'s are
practically assured in the large $\mu$ region if $A=0$ because then
they are lighter than the lightest neutralino.
\end{enumerate}

Since short lived gluinos ($\tau < 2 \times 10^{-11} \frac{m_{\gluino}}{1
GeV}$ sec) with masses between $\sim 4 - 126$ GeV are excluded
by missing energy searches (see ref. \cite{f:95} for discussion and
references), we can restrict the large $\mu$ parameter space for $A
\sim 1$ by requiring\cite{f:95}
\eq
\tau_{\gluino} = \frac{128 \pi cos^2 \theta_W}{\alpha_s \alpha_{em}}
f[\frac{m_{\chi}}{m_{\gluino}}] \frac{M_{sq}^4}{m_{\gluino}^5}>
2 \times 10^{-11} m_{\gluino} \frac{\rm sec}{\rm GeV},
\label{taugluino}
\en
where $f[y]$ is the phase space suppression when the lightest
neutralino mass is a non-negligible fraction of the gluino mass;
$f[0]=1$.  We replace $f\rightarrow 1$ to get a rough
estimate.  Then inequality (\ref{taugluino}) requires that either
$m_{\gluino} \gsi 126$ GeV or
\eq
m_{\gluino} \lsi 28~ {\rm GeV} \left(\frac{\mtilde}{10 \rm
TeV}\right)^{2/3} f^{1/6}.
\en
The latter condition requires that
\eq
A(1 + cot^2 \beta)~ log\left(\frac{M_{initial}^2}{m_W^2}\right) \lsi 6.6
\left(\frac{\mtilde}{10 \rm TeV}\right)^{-1/3} f^{1/6}
\en
be satisfied\footnote{The correction to these limits coming from
retaining the phase space factor $f$ is small unless $\mtilde
<< 10$ TeV or $\mu$ and $\mtilde$ are very large.  For instance, with
$m_{\gluino} \sim 28$ GeV and $m_{\chi} \sim 10$ GeV,
$f[\frac{10}{28}] = 0.56$ reducing the limit in comparison to the
rough estimate by $10\%$.}.

Fully studying the constraints on this scenario coming from requiring
relic particles not to overclose the universe is beyond the scope of
this paper.  In the usual scenario with $A \sim 1$, and tree level
gaugino masses taken to be proportional to the squark masses,
these considerations are used to rule out the existance of stable
neutralinos having mass less than a few GeV\footnote{See, e.g., ref.
\cite{drees,roszkowski}.}.  However the contribution of a relic to the
present mass density of the universe $\sim \frac{1}{<\sigma_{annih}
v>}$, and is therefore more weakly dependent on the relic mass than
on the squark mass because $\sigma_{annih} \sim M_{sq}^{-4}$.
Furthermore, when the gluino is light the availability
of the reaction $\chi \gluino \rightarrow q \bar{q}$ enhances
the annihilation of the neutralino, because the cross section is
larger by a factor $\sim \frac{\alpha_s}{\alpha_{em}}$ and because,
unlike $\chi \chi$ annihilation, it can go via the s-wave so the cross
section is non-vanishing in the non-relativistic limit\cite{f:98}.
Thus annihilation of neutralinos is more efficient in this
scenario even for the same squark mass and, more importantly, limiting
the squark mass puts different constraints on the gluino mass than in
the usual scenario.  For the large $\mu$ region this can be analysed
without difficulty.  However when the gluino
and photino are in the $\lsi 1$ GeV range, the freeze-out temperature
is of the same order of magnitude as the QCD confinement phase
transition temperature, so that the discussion of this scenario is
considerably more complicated than in the usual case and detailed
analysis is required to make quantitative statements.

Since a chargino has not been seen at LEP, we infered in Section
\ref{charged} that $\mu$, the supersymmetric coupling between the two
Higgs doublets, is either less than $\sim 100$ GeV or greater than
several TeV in this scenario.  If a chargino is not discovered at LEPII,
the low $\mu$ region would also be excluded.  If it were possible
to exclude the large $\mu$ region on other grounds, this would mean
that the present scenario could be definitively excluded at LEPII.
We have not made a comprehensive study of other constraints on $\mu$,
but note that for a given model of ew symmetry breaking only
certain regions of $\mu$ will be allowed.  For instance, the radiative
breaking scenario as discussed in refs. \cite{a-gpw,ir} does not work
in the large $\mu$ region when $A=0$.

\section{Summary}
\hspace*{2em}

We have investigated radiative corrections to gaugino masses,
revealing a number of interesting new possibilities for the
gaugino sector of a supersymmetrized standard model.  Constraining the
parameters of the model so that the lightest supersymmetric charged
particles are consistent with experimental bounds, we find that if
R-invariance is only broken spontaneously or if the dimension-3
SUSY-breaking parameters which explicitly violate R-invariance are
absent, the lightest neutralino is typically heavier than the gluino.
In the low $\mu$ region, the masses of the gluino and lightest
neutralino are less than $\sim 2$ GeV, even when $A$, the dimension-3
squark-squark-Higgs coupling, is non-zero.  In the large $\mu$ region
the lightest neutralino is heavier than $\sim$ 10 GeV and is more massive
than the gluino unless $A \ne 0$.  Thus the lightest gluino-containing
hadron naturally tends to be long-lived or even stable, and can be
consistent with laboratory searches\cite{f:95}. While this scenario is
very unusual from the phenomenological and cosmological points of
view, it may be consistent with observations. Further work is needed
to constrain the parameters of the model from considerations other
than just charged particle masses, and to explore the experimental and
cosmological implications of this scenario in greater
detail\footnote{A discussion of various astrophysical consequences of
a stable gluino can be found in a recent preprint by
Plaga\cite{plaga}.}. A more complete discussion of these issues is
left to the future\footnote{{\it Note Added:}  We wish to thank D.
Pierce for calling our attention to his paper with A.
Papadopoulos\cite{pierce_papa} which deals with some of the issues we
discuss here.  They also determine the radiative corrections to
chargino and neutralino masses.  In principle the case we treat should
be obtainable as a special case of their formulae, however our
expressions are considerably more compact and transparent than theirs
and we have not attempted to make a comparison.  Their work is
complementary to ours, in that it focuses on the possiblity of
extracting information on the GUT-scale mass relations from observed
sparticle masses, assuming general tree-level gaugino masses.  By
virtue of their interest in generality, they did not explore in detail
the scenario which we find most interesting, namely the possible
absence of dimension-3 susy-breaking terms.  The consequences of this
form of susy-breaking for the phenomenologically crucial issue of the
relative masses of gluino and lightest neutralino is the main new
feature of the present work.  Specializing to the portions of their
discussion relevant to a massless gaugino scenario, one finds that the
regions of parameter space considered acceptable in ref.
\cite{pierce_papa} differ from ours in important ways.  For instance
we find that for tree-level massless gauginos, the lower limit on the
chargino mass severly restricts the $\mu -  \beta$ space, so the large
values of $tan \beta$ which they consider (see e.g., their Figs. 2 and
4) are actually excluded in this scenario, at least as long as we are in
a region of parameter space where our global SU(2) ansatz holds.  Another
important difference is that in their discussion of the neutralino
sector, they seem to consider a tree-level neutralino mass to be a
necessity as a result of using LEP and CDF limits which are, however,
not applicable when one considers the very specific phenomenology
which follows from the absence of dimension-3 SUSY-breaking as
described here and in ref. \cite{f:95}.}.

{\bf Acknowledgements:}  We have had helpful conversations with many
colleagues, including T. Banks, R. Barbieri, L. Hall, E. Kolb, A.
Nelson, N. Seiberg, P. Sikivie, M. Strassler and J. Valle.


\newpage


\end{document}